\documentclass[epjST]{svjour}
\usepackage{graphics}
\begin{document}
\title{Finite length effect on supercurrents between trivial and topological superconductors}
\author{Jorge Cayao\inst{1}\fnmsep\thanks{\email{jorge.cayao@physics.uu.se}} \and Annica M. Black-Schaffer\inst{1}}
\institute{Department of Physics and Astronomy, Uppsala University, Box 516, S-751 20 Uppsala, Sweden}
\abstract{
We numerically analyze the effect of finite length of the superconducting regions on the low-energy spectrum, current-phase curves, and critical currents in junctions between trivial and topological superconductors. Such junctions are assumed to arise in nanowires with strong spin-orbit coupling under external magnetic fields and proximity-induced superconductivity. 
 We  show that all these quantities exhibit a strong dependence on the length of the topological sector in the topological phase and serve as indicators of the topological phase and thus the emergence of Majorana bound states at the end of the topological superconductor.
} 
\maketitle
\section{Introduction}
\label{intro}
The search for Majorana bound states (MBSs) in condensed matter physics has recently spurred a huge interest, further enhanced by its potential applications in topological quantum computation \cite{kitaev,RevModPhys.80.1083,Sarma:16,scharade18}.
In one dimension these exotic states emerge as zero-energy end states in topological superconducting nanowires, which can be achieved by combining common ingredients such as strong spin-orbit coupling (SOC), magnetic field, and proximity- induced conventional superconductivity \cite{PhysRevLett.105.077001,PhysRevLett.105.177002,Alicea:PRB10}.  

A quantized differential conductance with steps of height $2e^{2}/h$ at zero bias \cite{Law:PRL09} in normal-superconductor (NS) junctions is one of the most anticipated experimental signatures of MBSs and has motivated an enormous experimental effort since 2012 \cite{Mourik:S12,xu,Das:NP12,Finck:PRL13,Churchill:PRB13,Lee:13}, where initial difficulties
\cite{Lee:PRL12,Pientka:PRL12,Bagrets:PRL12,Liu:PRL12,Lee:13,Finck:PRL13,Churchill:PRB13,Rainis:PRB13,Sau:13,Zitkoetal} were  solved and high quality interfaces have recently been reported \cite{chang15,Higginbotham,Krogstrup15,zhang16,Albrecht16,Deng16,Nichele17,Suominen17,Marcus17,zhang18}. 
Despite all the efforts, there is however still controversy in the distinction between Andreev bound states and MBSs as in both cases similar conductance signatures might arise due to non-homogeneous chemical potentials \cite{PhysRevB.91.024514,StickDas17,Fer18}. It is therefore  important to go beyond zero-bias anomalies in NS junctions and study other geometries and signatures \cite{Prada:PRB17,Marcus17}. For recent reviews see Refs.\,\cite{Aguadoreview17,LutchynReview08}

One promising route includes  superconductor-normal-superconductor (SNS)  junctions based on nanowires which are predicted to exhibit a fractional $4\pi$-periodic Josephson effect in the presence of MBSs \cite{kitaev,Fu:PRB09,Kwon:EPJB03}, as a result of the protected fermionic parity as a function of the superconducting phase difference $\phi$ across the junction. 
Although the $4\pi$-periodic Josephson effect is difficult achieve as it disappears in thermal equilibrium,
 SNS junctions have  motivated both interesting theoretical studies \cite{PhysRevLett.103.107002,Badiane:PRL11,San-Jose:11a,PhysRevB.85.104514,Pikulin:PRB12,PhysRevB.86.140503,PhysRevB.87.104513,PhysRevB.88.144507,PhysRevB.91.024514,PhysRevB.92.134508,PhysRevB.93.220507,PhysRevB.94.085409,Hansen16,PhysRevB.94.205125,Cayao17b,PhysRevB.95.195430,PhysRevB.96.024516,PhysRevB.96.125438,PhysRevB.95.155449,Mirceacondmat17,Cayao18a,zazunov18,0953-8984-30-14-145402,PhysRevLett.120.267002} and promising experimental activity \cite{Rokhinson:NP12,Wiedenmann16,Deacon17,Bocquillon17,zuo17,Laroche18}. 
Another possibility recently considered for further evidence of MBSs is multiple Andreev reflection transport in voltage-biased SNS junctions 
\cite{SanJoseNJP:13,Kjaergaard17,Goffman17}. 
But even without additional voltages and at thermal equilibrium there exists proposals for signatures MBSs in SNS junctions. In particular, supercurrents in finite length SNS junctions, despite their overall 2$\pi$-periodicity, have very recently been reported to contain useful information about both the nontrivial topology and MBSs \cite{Cayao17b,Cayao18a}.

In this work we perform a numerical study of the low-energy spectrum, supercurrents, and critical currents in a trivial superconductor-topological superconductor junction  based in nanowires with strong SOC. Our work serves as a complementary study to previous reports where the left and right finite length S regions in  SNS junctions were both in the topological regime with four MBSs \cite{Cayao17b,Cayao18a}.  
We find that, unlike in fully topologically trivial junctions, in the topological phase the low-energy spectrum and current-phase curves are strongly dependent on the length of the topological S, which we can directly attribute to the emergence of MBSs and their hybridization. We also obtain that magnetic field dependence of the critical current is almost independent of the lengths of the superconducting regions in the trivial phase. However, in the topological phase the critical current develops oscillations with the magnetic field. These oscillations are connected to the emergence of MBSs but are reduced with increasing the length of the topological S region as the hybridization overlap of the MBSs is then strongly suppressed. We do not observe clear features of the topological transition point, an effect we mainly attribute  to the absence of Zeeman field in the left (trivial) region.

The remaining of this work is organized as follows. In Sec.\,\ref{sec1} we describe the model for SNS junctions based on nanowires with SOC. In Sec.\,\ref{sec2} we discuss the phase dependent low-energy spectrum and in Sec.\,\ref{sec3} we calculate and analyze the supercurrents as well as critical currents. In Sec.\,\ref{sec4} we present our conclusions.

\section{Model}
\label{sec1}
We consider a single channel nanowire with strong SOC and magnetic field modeled by \cite{rashba84a,rashba84b,PhysRevB.66.073311,PhysRevLett.90.256601,SOCExp10,Heedt17,PhysRevB.97.035433}
\begin{equation}
\label{H0Hamil}
H_{0}=\frac{p^{2}_{x}}{2m}-\mu-\frac{\alpha_{\rm R}}{\hbar}\sigma_{y}p_{x}+B\sigma_{x}\,,
\end{equation}
where $p_{x}=-i\hbar\partial_{x}$ is the momentum operator and $\mu$ the chemical potential, which determines the electron filling of the nanowire. Furthermore, $\alpha_{\rm R}$ represents the strength of Rashba SOC while $B=g\mu_{\rm B}\mathcal{B}/2$ is the Zeeman energy as a result of the applied magnetic field $\mathcal{B}$ in the $x$-direction along the wire, with $g$ being the wire $g$-factor and $\mu_{\rm B}$ the Bohr magneton. We use parameters for InSb nanowires, which include  the electron's effective mass $m=0.015m_{e}$, with $m_{e}$ the electron's mass, and the SOC strength $\alpha_{R}=50$\,meVnm  which is approximately 2.5 larger than the initial reported values \cite{Mourik:S12} and supported by recent experiments in InSb nanowires \cite{PhysRevB.91.201413,Kammhuber17}.

\begin{center}\begin{figure}
\resizebox{0.99\columnwidth}{!}{%
  \includegraphics{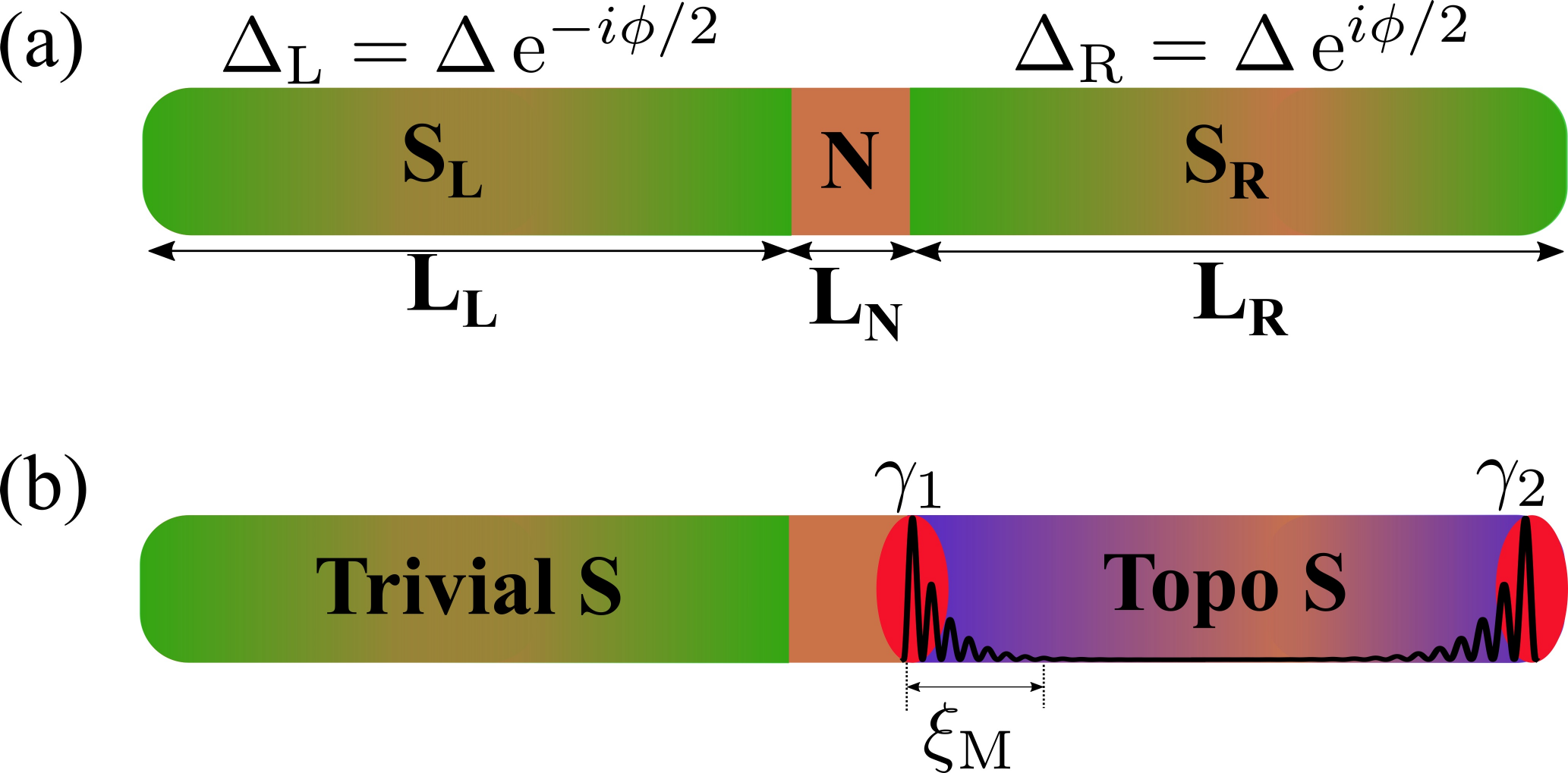} }
\caption{(a) The left and right regions of a nanowire with SOC are in contact with $s$-wave superconductors which induce superconducting correlations into the nanowire characterized by  pairing potentials $\Delta_{L,R}$, while the central region remains in its normal state. (b) A magnetic field applied solely to the right sector $S_{\rm R}$ drives it into the topological superconducting phase with Majorana bound states $\gamma_{1,2}$ at the ends with localization length $\xi_{\rm M}$.}
\label{fig:0}       
\end{figure}
\end{center}

For computational purposes, the model given by Eq.\,(\ref{H0Hamil}) is discretized on a tight-binding lattice such that $H_{0}=\sum_{i}c_{i}^{\dagger}hc_{i} + \sum_{\langle ij\rangle}c_{i}^{\dagger}vc_{j}+h.c.\,,$
where  $\langle ij \rangle$ denotes that $v$ couples nearest-neighbor $i,j$ sites. Here 
$h=(2t-\mu)\sigma_{0}+B\sigma_{x}$ and $v=-t\sigma_{0}+it_{\rm SO}\sigma_{y}$ are matrices in spin space, with $t=\hbar^{2}/(2m^{*}a^{2})$ being the hopping parameter, $t_{\rm SOC}=\alpha_{\rm R}/(2a)$ the SOC strengthm and $a$ the lattice spacing.  We consider a lattice spacing $a=10$\,nm.
Using open boundary condition the nanowire is automatically of finite length. We then assume that the left and right sections of the nanowire are in close proximity to $s$-wave superconductors. This induces finite superconducting pairing correlations into the nanowire characterized by the mean-field order parameter $\Delta_{L,R}=\Delta{\rm e}^{\pm i\phi/2}$, where $\phi$ is the superconducting phase difference across the junction, while the middle region remains in the normal state.  This leads to a SNS junction, where the 
left S, normal N, and right S regions are of finite length $L_{\rm L}$, $L_{\rm N}$ and  $L_{\rm R}$, respectively, as schematically shown in Fig.\,\ref{fig:0}(a). We here consider very short junctions, such that $L_{\rm N} = 20$\,nm, and keep same chemical potential $\mu$ in all three regions for simplicity. We consider the total lengths of the wire ($L_{\rm L}+L_{\rm R}+L_{\rm N}$)  to be between 600\,nm to 2300\,nm, consistent with typical lengths in experiments \cite{Mourik:S12,chang15,zuo17,zhang16,gul17}; these values correspond to 60 and 230 lattice sites, respectively,  with a lattice spacing of $a=10$\,nm in our simulations. The effective junction is thus set by the finite phase difference between left and right S regions.
The numerical treatment of the superconducting correlations are carried out within the standard Nambu representation, see 
Refs.\cite{Cayao17b,Cayao18a}.
Furthermore, we assume that the magnetic field $\mathcal{B}$ is applied solely to the right S region of the nanowire, which can be achieved e.g.~by contacting the right S to a ferromagnetic material \cite{Eschrig2007,7870d3ff91ed485fa3e55e901ff81c80,LinderNat15}.
The left S and N regions are not subjected to any magnetic field. This allows us to drive the right S region into the topological phase with MBSs $\gamma_{1,2}$ at both its ends for $B>B_{c}$ as depicted in Fig.\,\ref{fig:0}(b), where $B_{c}=\sqrt{\mu^{2}+\Delta^{2}}$ is the critical field \cite{PhysRevLett.105.077001,PhysRevLett.105.177002,Alicea:PRB10}. 
For $B<B_{c}$ the whole system is thus topologically trivial and no MBSs are expected. The MBSs in the right S region are localized to its two end points and decay exponentially into the middle of the S region in an oscillatory fashion with a decay length $\xi_{\rm M}$, developing an spatial overlap due to the finite length $L_{\rm R}$ when $L_{\rm R}\leq 2\xi_{\rm M}$ \cite{PhysRevB.86.180503,DasSarma:PRB12,PhysRevB.86.085408,Rainis:PRB13}. It is worth pointing out that we have verified (not shown) that the  wavefunction associated to $\gamma_{1}$ has a small non-oscillatory tail that decays into  N and also slightly leaks into the left S region. In contrast, for long N regions with finite magnetic field, the wavefunction exhibits an oscillatory behavior that does not decay \cite{PhysRevB.86.085408,Cayao18a}.

Using this model we perform numerical diagonalization to investigate the low-energy spectrum and supercurrents across many different SNS junctions, in particular varying size of the two S regions, as well as superconducting phase and magnetic field strength.

\begin{center}\begin{figure}
\resizebox{0.99\columnwidth}{!}{%
  \includegraphics{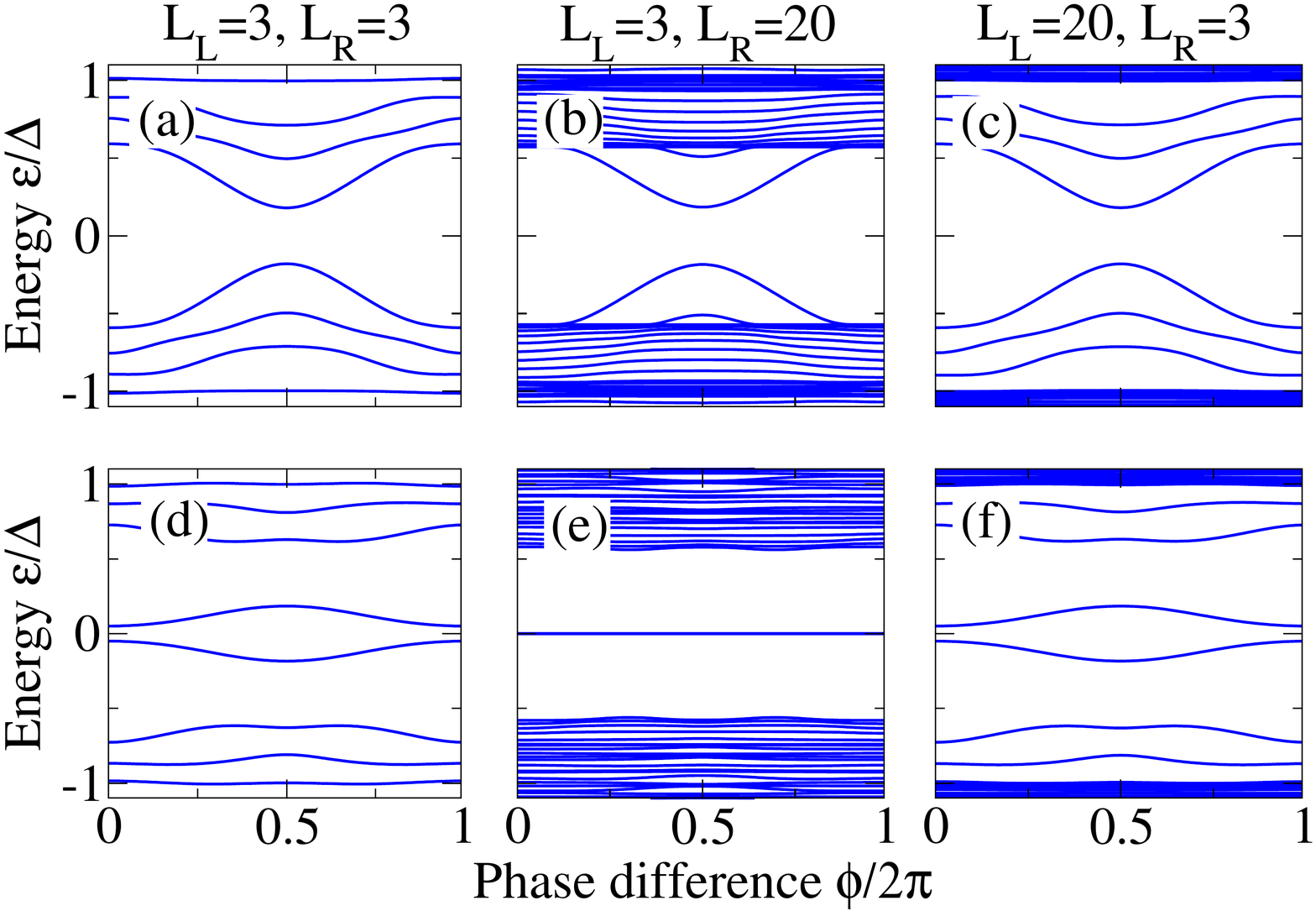} }
\caption{Phase dependent low-energy spectrum in the trivial phase $B=0.5B_{c}$ (top row) and topological phase at $B=1.5B_{c}$ (bottom row). Different panels correspond to different values of the length of the left ($L_{\rm L}$) and right  ($L_{\rm R}$) S regions.
Notably, an increase in $L_{\rm R}$ changes the low-energy levels in the topological phase (e), but does not affect the trivial phase (b). Lengths are given in units of $100$\,nm. Parameters: $\Delta=0.9$meV, $\alpha_{\rm R}=50$meVnm, $\mu_{L,R}=0.5$meV.}
\label{fig:1}       
\end{figure}
\end{center}

\begin{center}\begin{figure}
\resizebox{0.99\columnwidth}{!}{%
  \includegraphics{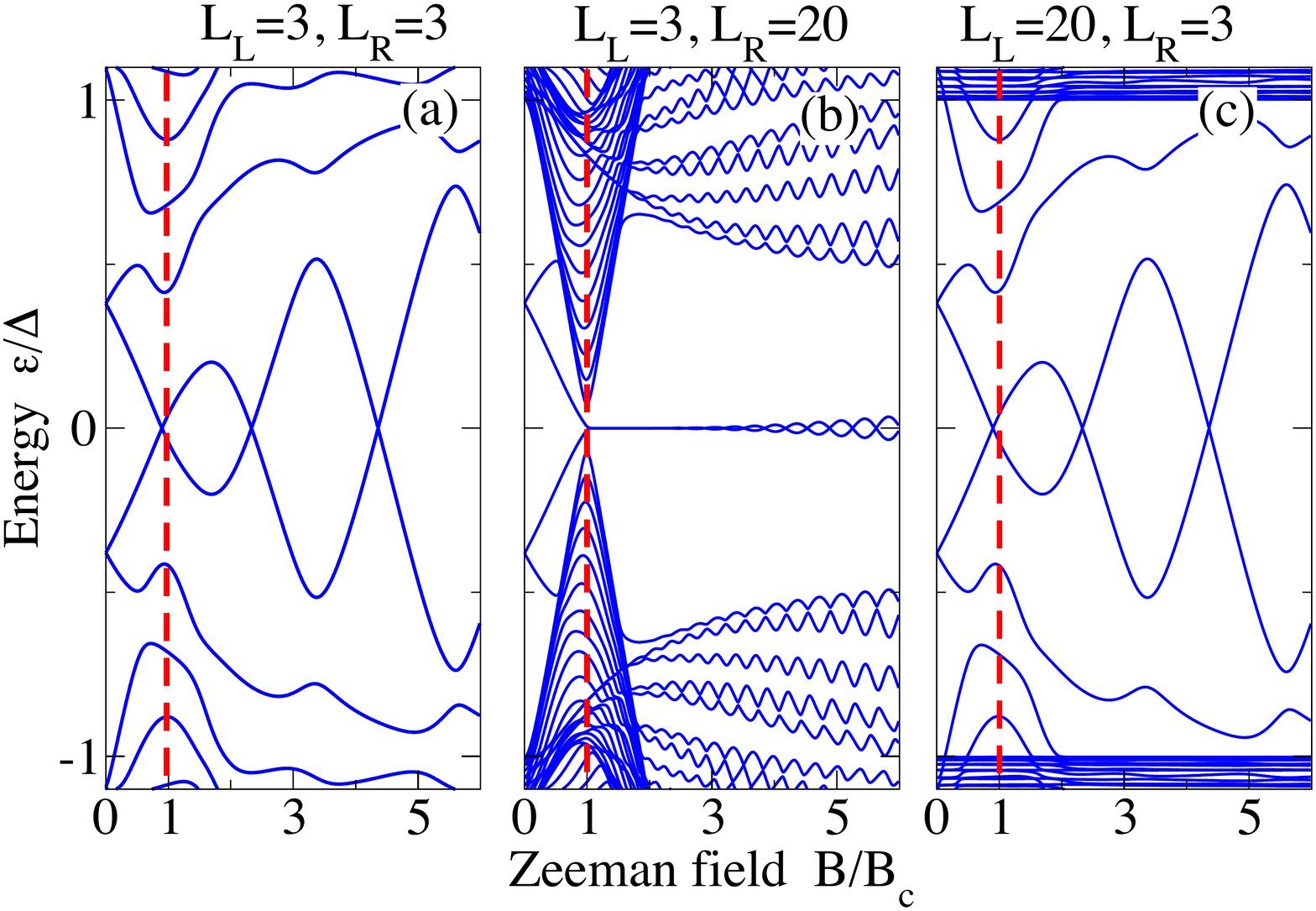} }
\caption{Magnetic field dependent low-energy spectrum at $\phi=\pi$  for equal S region lengths (a), and larger (b) and shorter (c)  right region lengths. Topological phase transition $B=B_{c}$ is indicated by vertical dashed red lines. Notably, the low-energy spectrum is solely affected  by changes in the length of the right S sector. 
 Lengths are given in units of $100$\,nm. Parameters: $\Delta=0.9$meV, $\alpha_{\rm R}=50$meVnm, $\mu_{L,R}=0.5$meV.}
\label{fig:2}       
\end{figure}
\end{center}

\section{Energy spectrum}
\label{sec2}
In this section we investigate the evolution of the low-lying energy levels $\varepsilon_{p}$ in a short SNS junction under a magnetic field applied to the right S region and for different superconducting phases $\phi$. 

Due to the finite length of the whole SNS structure, the energy spectrum is discrete and Andreev reflections at the junction interface together with a finite superconducting phase difference lead to the formation of Andreev bound states within the energy gap $\Delta$. In very short junctions the spin-orbit coupling does not split the energy levels \cite{PhysRevB.77.045311,Dimitrova2006,PhysRevB.77.045311,SanJoseNJP:13,PhysRevB.96.075404,PhysRevB.96.014519,Cayao18a}  but a Zeeman field generally does. Most importantly, the low-energy spectrum acquires a phase dependence that allows the identification of MBSs in the topological phase \cite{Cayao17b,Cayao18a}.

In Fig. \,\ref{fig:1} we show the phase dependent low-energy spectrum for different values of the length of the left  ($L_{\rm L}$) and right ($L_{\rm R}$) S  regions in the trivial $B<B_{c}$ (top row) and topological $B>B_{c}$ (bottom row) phases. In the case of equal and short S region lengths (a, d) the low-energy spectrum is very sparse and exhibits an appreciable phase dependence with a marked difference between the trivial (top row) and topological phase (bottom row). In the trivial phase $B<B_{c}$, the low-energy levels behave as conventional Andreev states which tend towards zero energy at $\phi=\pi$ \cite{Cayao18a}, as seen in Fig. \,\ref{fig:1}(a). However, unlike predicted by the standard theory  \cite{Beenakker:92}, the minimum energy at $\phi=\pi$ is here non zero mainly because our junction is away from the Andreev approximation where the chemical potential $\mu$ is assumed to be the dominating energy scale \cite{Cayao18a}. The situation is distinctly different in the topological phase in Fig. \,\ref{fig:1}(d), where two levels emerge around zero energy with an energy splitting for all phase differences $\phi$, which becomes largest at $\phi=\pi$. These are the two MBSs formed at either of the topological S$_{\rm R}$ region. It is the finite overlap of the two MBSs across the $S_{\rm R}$ region that causes the energy splitting away from zero. 
Indeed, as the length of the topological S$_{\rm R}$ region is increased, the splitting of MBSs is exponentially reduced such it even completely vanishes for very long regions as seen in Fig.~\ref{fig:1}(e), where the MBSs acquire their zero-energy character irrespective of the phase difference. The increase of $L_{\rm R}$  also introduces more energy levels to the quasicontinuum (dense set of levels above the minigap in (b,c,e,f)), but it notably does not modify the low-energy behavior.  
Further evidence that the energy splitting in (d) is due to MBSs is acquired by instead increasing the length of the left region ($L_{\rm L}$), as done in Fig.~\ref{fig:1}(c,f). In this case the low-energy spectrum is not altered with respect to the case with equal lengths (a,d).
 
Additional and complementary information is given by the magnetic field dependence of the low-energy spectrum, which we present in Fig.\,\ref{fig:2} for $\phi=\pi$ and different values of $L_{\rm L(R)}$. We directly notice how the the magnetic field dependent low-energy spectrum captures the gap closing and the emergence of MBSs for $B>B_{c}$, as well as the MBS hybridization through the oscillatory energy levels around zero energy for $B>B_{c}$. The gap closing is here not sharp primarily due to the finite length of the system and but also due to relatively large values of the SOC. Although the SOC does not determine the critical $B_{c}$, it does affects the sharpness of the gap closing in finite length systems. We also clearly see that the MBSs energy splitting is significantly reduced by increasing the length of the topological  S$_{\rm R}$ region. However, a similar increase in the length of the trivial left region does not introduce any change in the low-energy spectrum, but only give rise to a dense set of levels around $\Delta$, as seen in Fig.\,\ref{fig:2}(c).

\section{Supercurrents and critical currents}
\label{sec3}%
After the above discussion on the low-energy spectrum we now investigate the supercurrents in the SNS junctions,  which can be directly calculated  from the discrete Andreev spectrum 
$\varepsilon_{p}$ as \cite{Beenakker:92,Cayao17b}
\begin{equation}
\label{josehpcurrent}
I(\phi)=-\frac{e}{\hbar}\sum_{p>0}{\rm tanh}\Big(\frac{\varepsilon_{p}}{2\kappa_{B}^{}T} \Big)\frac{d\varepsilon_{p}}{d\phi}\,,
\end{equation}
where $\kappa_{B}$ is the Boltzmann constant, $T$ the temperature, and the summation is performed over all positive eigenvalues $\varepsilon_{p}$ of the Hamiltonian described our SNS junction.  Previous equation is valid for finite length SNS junctions and, in principle, for short and long junctions. Here we only discuss  short SNS junctions. For long junctions, see \cite{Cayao18a}.

\begin{center}\begin{figure}
\resizebox{0.99\columnwidth}{!}{%
  \includegraphics{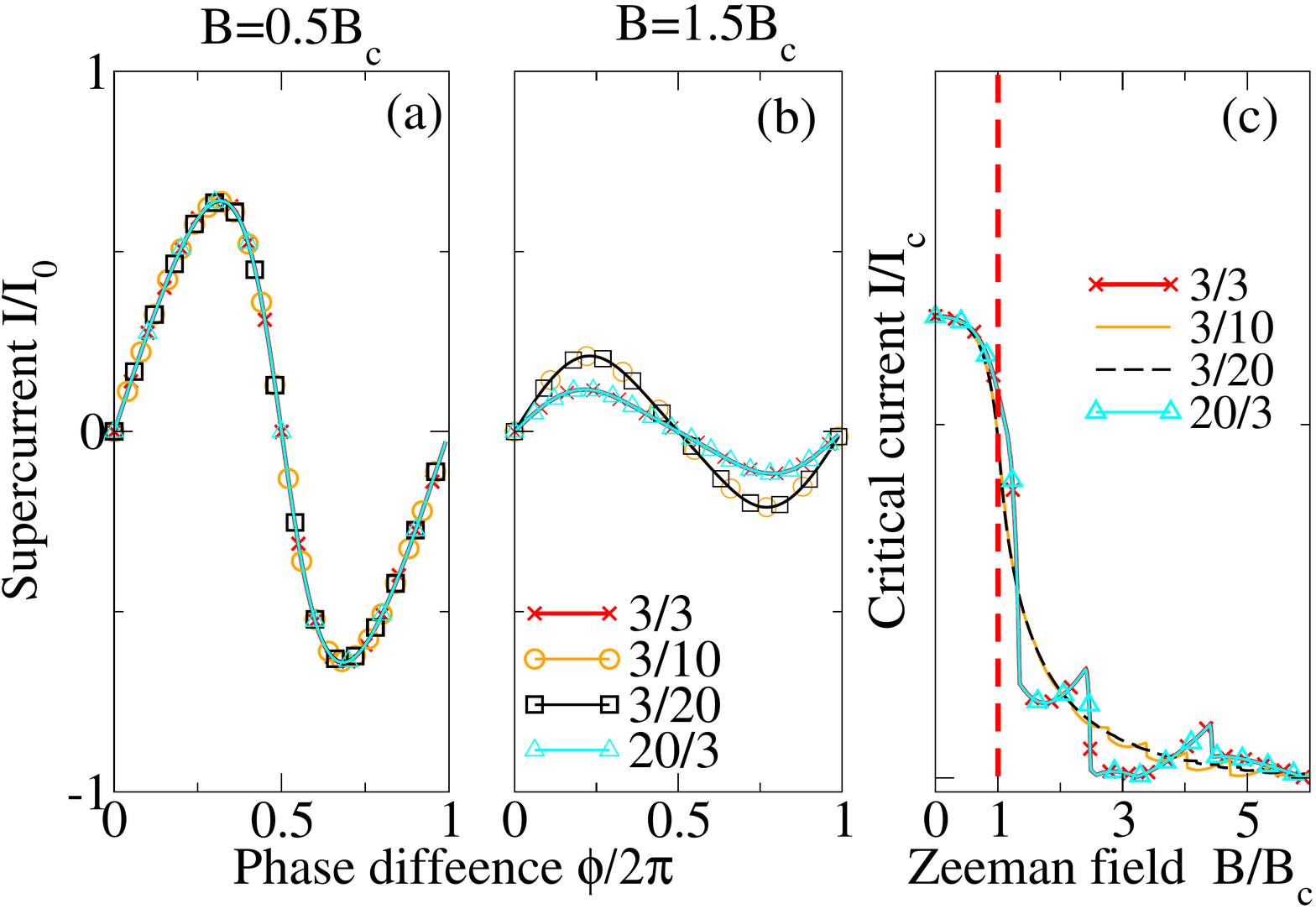} }
\caption{Phase dependent supercurrents in the trivial $B=0.5B_{c}$ (a) and topological phase at $B=1.5B_{c}$ (b), as well as and magnetic field dependent critical currents (c). Different curves correspond to different values of $L_{\rm R,L}$.
 Notably, an increase in $L_{\rm R}$ affects the supercurrent and critical currents in the topological phase (b,c), but does not affect the trivial phase (a,c). Lengths are given in units of $100$\,nm. Parameters: $\Delta=0.9$meV, $\alpha_{\rm R}=50$meVnm, $\mu_{L,R}=0.5$meV.}
\label{fig:3}       
\end{figure}
\end{center}

In Fig.\,(\ref{fig:3})(a,b) we plot the phase dependence of th supercurrents $I(\phi)$ in the trivial $B=0.5B_{c}$ (a) and topological $B=1.5B_{c}$ (b) phases at $T=0$. A general feature is that in both the trivial and topological phases the supercurrents are $2\pi$-periodic $I(\phi)=I(\phi+2\pi)$ \cite{Cayao17b,PhysRevB.96.024516,Cayao18a,zazunov18}, developing its maximum value close to $\phi=\pi/2$. This is in contrast to the case considering purely semi-infinite topological junctions which report $4\pi$-periodicity of $I(\phi)$ when the system is not allowed to relax to thermal equilibrium for each phase \cite{kitaev,Fu:PRB09,Kwon:EPJB03}.  In the trivial phase supercurrents acquire a sine-like behavior and do not exhibit any change upon variations of the lengths of either right or left S region, as can be seen in Fig.~\ref{fig:3}(a). Changing the magnetic field from $B=0.5B_{c}$, but still within the trivial phase, do not alter the magnitude of $I(\phi)$ but only introduce a small zig-zag feature around $\phi=\pi$, similar to Ref.~\cite{Cayao18a}. 

In the topological phase the overall magnitude of the supercurrents $I(\phi)$ are reduced due to the higher magnetic fields, as seen in Fig.\,(\ref{fig:3})(b). More importantly though,  $I(\phi)$  undergoes an enhancement due to the reduction of the MBSs energy splitting  when $L_{\rm R}$ increases. We have verified that for very large magnetic fields $B\gg B_{c}$, the overall supercurrent is reduced and eventually completely suppressed due to the different pairing symmetries in the trivial  and topological  superconducting regions \cite{PhysRevB.93.125435,PhysRevB.96.024516}.  
This can be understood as follows: the superconducting correlations in the left S have spin singlet $s$-wave symmetry and also mixed spin triplet $p$-wave  due to finite SOC  both with $m_{z}=0$ for the Cooper pairs \cite{PhysRevB.92.134512,cayao18c}, while in the right S there is a coexistence of correlations with spin-singlet ($m_{z}=0$)  $s$-wave, equal spin-triplet ($m_{z}=\pm1$) $p$-wave, and mixed spin-triplet ($m_{z}=0$) $p$-wave due to the finite magnetic field in such region. 
For large values of Zeeman fields, the spin-singlet $s$-wave and mixed spin-triplet $p$-wave correlations both with $m_{z}=0$ are suppressed due to Zeeman depairing, leaving only equal spin-triplet ($m_{z}=\pm1$) $p$-wave correlations in the right S region. 
In total, this give rise to an incompatibility between the  trivial region S with spin-singlet and mixed spin-triplet ($m_{z}=0$) correlations
 and the fully equal spin-triplet ($m_{z}=\pm1$) $p$-wave state in the topological region at extremely large magnetic fields, thus reducing the supercurrent. Still, there is a significant region for $B>B_{c}$ where the enhancement observed in Fig.\,(\ref{fig:3})(b) in the supercurrent is still useful to identify the topological phase and its MBSs.

Further signatures of MBSs can be acquired from the critical currents $I_{\rm c}$, which is simply the  maximum supercurrent that  flows across the junction, which can be calculated  by maximizing the  supercurrent $I(\phi)$ with respect to the superconducting 
phase difference $\phi$, 
\begin{equation}
I_{\rm c}={\rm max_{\phi}}[I(\phi)]\,,
\end{equation}
where $I(\phi)$ is numerically found using Eq.\,(\ref{josehpcurrent}). In what follows we solely discuss the zero temperature situation $T=0$.

In Fig.\,(\ref{fig:3})(c) we present the magnetic field dependence of the critical currents for different values of the lengths of the left $L_{\rm L}$ and right $L_{\rm R}$ S regions. At $B=0$ the critical current is finite and maximum, while it decreases as the magnetic field increases. In the trivial phase, for$B<B_{c}$, the critical currents are exceptionally independent of variations of $L_{\rm {L,R}}$, as seen in Fig.\,(\ref{fig:3})(c). A very different behavior is observed in the topological phase. First, at $B=B_{c}$, $I_{c}$ is still finite, however, the kink-like feature reported in some previous studies \cite{Cayao17b,Cayao18a,zazunov18} is absent mainly due to the zero Zeeman field $B$ in the left region. 
We have verified that the SOC, finite length of the S regions, and induced gap also affect the visibility of such kink at $B_{c}$ but it is more detrimental the absence of $B$ in the left region.
 Beyond the topological transition, for $B>B_{c}$, the critical current is further reduced and finally vanishes for extremely large magnetic fields, as is seen in  Fig.\,(\ref{fig:3})(c). This is due to the incompatibility between the superconducting correlation symmetries in the trivial region and in the topological region at extremely large magnetic fields, as explained before \cite{PhysRevB.93.125435,PhysRevB.96.024516}. However, much before that, the critical current captures the splitting of MBSs through  noticeable oscillations as function of the magnetic field. The oscillations are reduced when the length of the topological sector (right S) increases, an effect purely related to the energy splitting observed in the Zeeman dependent low-energy spectrum in Fig.\,\ref{fig:1}. We can therefore attribute this behavior to the emergence and subsequent hybridization of the MBSs, similar to previous reports when the left and right sectors become topological with four MBSs \cite{Cayao17b,Cayao18a}. Taken together, this introduces a strong dependence of the critical current on the length of the topological sector in the topological phase $B>B_{c}$, unlike in the trivial phase $B<B_{c}$ where critical currents are length independent. The length dependence of one of the two S regions can thus be used to determine the topological phase transition.

\section{Conclusions}
\label{sec4}
In this work we investigated the finite length effect of the superconducting sectors on the low-energy spectrum, supercurrents, and critical currents in junctions between trivial and topological superconductors based on nanowires with strong Rashba SOC. 
We demonstrated that the low-energy spectrum and current-phase relationship in the topological phase are strongly dependent on variations of the length of the topological S region, but show not dependence on length for the trivial S region. This effect we were able to trace back to the emergence of MBSs.
We also showed that the critical current reveal important information in the distinction between trivial and topological phases and thus offer a straightforward experimental signature for nontrivial topology. In particular, the critical current is essentially completely independent of the length of the superconducting regions in the trivial phase. However, in the topological phase there is both a length dependence and the critical current starts to exhibit notable oscillations, which are reduced as the length of the topological sector increases. The oscillations we were able to attribute to the MBSs at either end point of the topological S and their mutual hybridization. Thus both the current-phase relationship and critical current exhibit features that uniquely identifies the topological phase transition in SNS nanowire junctions. Notably, and in contrast to the elusive $4\pi$ fractional Josephson effect, both of these effects are accessible through very standard measurements and thus offers straightforward yet powerful signatures of nontrivial topology and MBSs.

\section*{Acknowledgments}
We thank E.~J.~H. Lee for interesting discussions.  This work was supported by the Swedish Research Council (Vetenskapsr\aa det, Grant No.~621-2014-3721), the G\"{o}ran Gustafsson Foundation, the Swedish Foundation for Strategic Research (SSF), and the Knut and Alice Wallenberg Foundation through the Wallenberg Academy Fellows program.

\bibliographystyle{epj}

\bibliography{biblio}

\end{document}